\documentclass[reviewcopy]{elsarticle}

\usepackage[reviewcopy]{adndt}

\usepackage{amsmath}    
\usepackage{amssymb}    
\usepackage{mathrsfs}   
\usepackage{longtable}  
\usepackage{aas_macros} 
\usepackage{graphicx}   
\usepackage{xcolor}     

\graphicspath{{./}{figs/}}

\biboptions{square,sort&compress}
\bibpunct[]{[}{]}{,}{n}{}{;}
\newcommand{\bmd}{$\beta^{-}$-decay}
\newcommand{\bdne}{$\beta^{-}$-delayed neutron emission}
\newcommand{\bd}{$\beta$-decay}
\newcommand{\bstab}{$\beta$-stable}
\newcommand{\bstr}{$\beta$-strength}
\newcommand{\qrpahf}{QRPA+HF}
\newcommand{\gray}{$\gamma$-ray}
\newcommand{\grays}{$\gamma$-rays}

\begin{document}

\begin{frontmatter}

\journal{Atomic Data and Nuclear Data Tables}


\title{Nuclear \bmd{} with statistical de-excitation}

\author[LANL,CTA]{M.~R. Mumpower\corref{cor1}}
\ead{E-mail: mumpower@lanl.gov}
  
\author[LANL]{T. Kawano}

\author[LANL]{O. Korobkin}

\author[LANLX]{G. W. Misch}

\author[LANL]{T. M. Sprouse}

\cortext[cor1]{Corresponding author.}

\address[LANL]{Theoretical Division, Los Alamos National Laboratory, Los Alamos, NM 87545, USA}
\address[LANLX]{Theoretical Design, Los Alamos National Laboratory, Los Alamos, NM 87545, USA}
\address[CTA]{Center for Theoretical Astrophysics, Los Alamos National Laboratory, Los Alamos, NM 87545, USA}

\date{10.29.2024} 

\begin{abstract}
The accurate description of nuclear \bmd{} has far-reaching consequences for applications spanning nuclear reactors to the creation of heavy elements in astrophysical environments. 
We present the nuclear particle spectra associated with the \bd{} of neutron-rich nuclei calculated with the well benchmarked coupled Quasi-particle Random Phase Approximation and Hauser-Feshbach (\qrpahf{}) model. 
This approach begins with the population of the daughter nucleus via semi-microscopic Gamow-Teller or First-Forbidden strength distributions (QRPA) and follows the statistical de-excitation (HF) until the initial available excitation energy is exhausted. 
At each stage of de-excitation the emission by neutrons and \grays{} is considered obeying quantum mechanical selection rules. 
For completeness we also provide parsed Auger and Internal Conversion (IC) electron spectra from Evaluated Nuclear Data Files (ENDF). 
Our results are tabulated and provided in parsable ASCII formatted tables that are suitable for inclusion in various applications. 
\end{abstract}

\end{frontmatter}

\newpage

\tableofcontents
\listofDtables
\listofDfigures
\vskip5pc


\section{Introduction} \label{sec:intro}
The weak interaction is responsible for the process of nuclear \bmd{} in which a neutron is converted into a proton with emission of an electron and electron anti-neutrino. 
In nuclei, this process transforms the parent nucleus ($Z$,$A$), with $Z$ protons and $A$ nucleons into a daughter nucleus, ($Z+1$, $A$), often leaving the daughter nucleus in an excited state that may decay via emission of \grays{}. 
If the precursor nucleus is sufficiently neutron-rich, the excitation energy may be above the neutron separation energy opening the channel for neutron emission. 
When there are sufficient levels in the daughter nucleus, a statistical picture of the subsequent nuclear de-excitation may be employed \cite{LaRana1987, Cole2000, Kawano2008, Mumpower2016, Mumpower2018, Mumpower2022}. 
The participating particles may be emitted at a host of energies and depend on the details of nuclear structure \cite{Algora2021}. 
An accurate description of the particle spectra and branching ratios that arise during this phenomenon are important inputs for a number of applications including the study of fission \cite{Zeynalov2009, Becker2013, Talou2018, Siegl2018b}, nuclear reactors \cite{Tobias1980, Algora2010, Wood2013} and associated reactor neutrinos \cite{Hayes2018, Hayes2021, Letourneau2023, Stereo2023}, tests of fundamental symmetries \cite{Barabash2023, Fang2023, Tikhonov2023}, and during the production of elements in astrophysical events \cite{Lattimer1974, Mumpower2012b, Mumpower2014, Horowitz2019} that may produce observable electromagnetic signatures \cite{Li1998, Martin2015, Tanvir2017, Wollaeger2017, Zhu2018, Korobkin2020, Fujimoto2020}. 

Experimental measurements of \bmd{} and \bdne{} have been performed by a number of groups \cite{Dillmann2014, Siegl2018a, Spyrou2016, Liddick2016, CaballeroFolch2016, Dillmann2018}. 
More recently, the population of $^{131}$Sn from $\beta$-decay of isomerically pure $^{131}$In has allowed for the identification of the delayed neutron emission probabilities from each isomeric state in $^{131}$In \cite{Benito2024}. 
In similar work, the $\beta$-delayed neutron spectroscopy of $^{70}$Co and $^{72}$Co was explored where the $\beta$-decay strength above the neutron separation energy of $^{70}$Ni was identified \cite{Siegl2024}. 
In the work of Yokoyama \textit{et al.} $\beta$-delayed 2 neutron emitters were studied in the Ga ($Z=31$) isotopic chain \cite{Yokoyama2023}. 
It was shown that the probability to emit neutrons in statistical model calculations are sensitive to the nuclear level densities, especially to the level density of the 1-neutron emitting daughter. 
Experimental evidence has also arisen for non-statistical decay, which may require a modification of the compound nucleus assumption in neutron-rich nuclei \cite{Heideman2023}. 
Using experimental data along the potassium isotopic chain, it has been proposed that non-statistical neutron emission may proceed via coupling with nearby doorway states that have sufficient neutron-emission probabilities \cite{Xu2024}. 
While other methods are being developed \cite{Gorton2023}, from a pragmatic perspective, the statistical picture of nuclear de-excitation is presently the only viable predictive model for large scale studies of the $\beta$-decay of atomic nuclei. 
Indeed, statistical de-excitation models have continually outperformed cut-off methods \cite{Moller2003} for predicting delayed neutron emission \cite{Dimitriou2021, Phong2022}. 

The distribution of excited states populated in the daughter nucleus, particularly in the continuum, is not always measured experimentally. 
This quantity can be estimated via techniques like the Quasi-particle Random Phase Approximation (QRPA). 
In analogy to other uses of strength functions in nuclear physics, the \bstr{} function refers to the behavior of the squares of overlap integrals. 
This quantity (definition provided in the next section) is typically given as an average over energy levels and may also include sums over multipolarity, parity, and spin of the final state \cite{Hansen1973}. 
Theoretical calculations of particle spectra and branching ratios have been performed in the past using simple models of the \bstr{} function \cite{Pappas1972, Shalev1974, Rudstam1974, Gjotterud1978, Mann1982}. 
More recent work on \bmd{} has been performed using microscopic approaches \cite{Borzov2014, Marketin2016} and with the finite amplitude method \cite{Mustonen2014,Shafer2016,Ney2020}, however, these works have not yet provided particle spectra. 

In this work we present the particle spectra associated with nuclear \bmd{} using the Los Alamos statistical de-excitation framework ~\cite{Kawano2008, Mumpower2016, Moller2019, Mumpower2022}. 
While the details of our model have been stated in previous work, we provide a brief overview of the assumptions in Sec.~\ref{sec:model}. 
In Section \ref{sec:pspec} we discuss the calculation of the particle spectra. 
In Section \ref{sec:output} we discuss the outputs of the model and end with a presentation of the results across the chart of nuclides which are also provided as supplementary ASCII formatted files. 

\section{The \qrpahf{} Model} \label{sec:model}
We employ the coupled Quasi-particle Random Phase Approximation and Hauser-Feshbach (\qrpahf{}) model in this work. 
This model comprises a two step process which first describes the production of the daughter nucleus via Quasi-particle Random Phase Approximation (QRPA) and then follows the statistical de-excitation that ensues via Hauser-Feshbach (HF) theory. 
In addition to \bmd{}, our model is also capable of describing prompt fission spectra as discussed in Refs.~\cite{Okumura2018, Jaffke2018}, but we do not cover this particular application here. 
We review the basic tenets and assumptions of our framework below. 

The initial population of the daughter from the parent is calculated using the QRPA theory of M\"{o}ller et al. \cite{Krumlinde1984, Moller1990, Moller1997, Moller2003}. 
In this model, the Schr\"{o}dinger equation is solved for the nuclear wave functions and single-particle energies. 
A folded-Yukawa Hamiltonian is used that describes finite density diffuseness \cite{Moller1981, Moller1995, Moller2018}. 
Invoking Fermi's Golden Rule, the rate to undergo $\beta$-decay from initial parent state, $\vert i \rangle$, to final daughter state, $\vert f \rangle$, is
\begin{equation}
    \lambda_{if} = \frac{2\pi}{\hbar}\vert \langle f \vert \widehat{O}_\beta \vert i \rangle \vert^2 \rho(E_f) \ ,
    \label{eq:fermi_gr}
\end{equation}
where $\langle f|\widehat{O}_\beta| i\rangle$ is the matrix element from initial to final state, and $\rho(E_f)$ is the density of final states. 
The decay matrix element, often written $M_{fi}$, depends on the details of the nuclear interaction for which $\widehat{O_\beta}$ may be replaced by the appropriate operator. 
Both Gamow-Teller and First-Forbidden transitions what about Fermi transitions? are considered as detailed in Refs.~\cite{Moller1997, Moller2003}. 
For this work we use the latest model of Ref.~\cite{Moller2019}. 

The \bstr{} function is defined as
\begin{equation}
   S_\beta(E_x) = \frac{I(E_x)}{F(Z+1,A,Q_{\beta}-E_x)T_{1/2}} \approx C \times \frac{\textrm{d}\overline{\sum{\vert M_{fi} \vert^2}}}{\textrm{d} E_x} \ , 
   \label{eq:S_beta}
\end{equation}
where $I(E_x)$ is the $\beta$-intensity per MeV of levels at energy $E_x$ in the daughter nucleus, $F$ is the Fermi function and $T_{1/2}$ is the half-life of the parent nucleus in seconds, $\overline{\sum{\vert M_{fi} \vert^2}}$ represents the smoothed sum of all allowed and forbidden matrix elements and $C \sim 6220$ is a collection of constants \cite{Duke1970, Hansen1973}. 
The units of $S_\beta$ are [s$^{-1} \times $MeV$^{-1}$] and the Fermi function for \bmd{} is defined as,
\begin{equation}
\begin{array}{cclr}
    F(Z,A,w) &=& 2(1+s)(2pR)^{2(s-1)}e^{\pi\eta}\left|\frac{\Gamma(s+i\eta)}{\Gamma(2s+1)}\right|^2 & \ , \\
    \textrm{with} \\
    s &=& \left[1-(\alpha Z)^2\right]^{1/2} & \ , \\
    \eta &=& \alpha Z \frac{w}{p} & \ , \\
    R &=& 2.908\times 10^{-3}A^{1/3} - 2.437A^{-1/3} & \ , 
    \label{eq:fermi_fnc}
\end{array}
\end{equation}
where $p$ is the final momentum, $R$ is the nuclear radius in electron Compton wavelengths \cite{Gove1971}, $Z$ and $A$ are the charge and nucleon number of the daughter nucleus, $w$ is the energy and $\Gamma$ is the Gamma function. 

The details of the \bstr{} distribution for the daughter nucleus are dependent on the level structure of this system. 
Experimental kowledge of this structure is often limited to a few levels or may be completely uncertain for neutron-rich nuclei \cite{Brown2018}. 
In the case where the level structure and \bd{} scheme are known completely, data may be incorporated from the Reference Input Parameter Library (RIPL-3) \cite{Capote2009} or Evaluated Nuclear Structure Data Files (ENSDF) and no QRPA strength is used. 
If the database is incomplete, we combine and renormalize the QRPA solutions with the ENSDF data as in Ref.~\cite{Mumpower2016}. 

Naturally, there is an intrinsic uncertainty associated with our theoretical QRPA modeling. 
For these reasons, we smooth the $\beta$-strength distribution by a Gaussian,
\begin{equation}
  \omega(E_x) = C_\omega \sum_k b^{(k)}
     \frac{1}{\sqrt{2\pi} \Omega}
     \exp\left\{
            - {\frac{ [E^{(k)} - E_x]^2 }{2\Omega^2}}
         \right\} \ ,
  \label{eq:omega}
\end{equation}
where $b^{(k)}$ are the branching ratios (the intensities of Eqn.~\ref{eq:S_beta}) from the parent state to the daughter states $E^{(k)}$, $E_x$ is the excitation energy of the daughter nucleus. 
The Gaussian width, $\Omega$, is taken to be a function of the mass number of the parent system, $\Omega(A)=8.62 / A^{0.57}$ keV. 
We have tried numerous other functions for this variable, including the constant $\Omega=100$ keV, as in Refs.~\cite{Kawano2008, Mumpower2016}, which does give some small variability in our results. 
We summarize the variation of our results based on three calculations of $\Omega \in \{30.0, 100.0, \Omega(A)\}$ in Table \ref{table:dev}. 
The predicted particle multiplicity is relatively insensitive to the choice of $\Omega$ while the average maximum energy change between the sample calculations is between $100$ to $500$ keV. 
The deviation in electron and neutrino average energies more strongly depend on the choice of this parameter than the average \gray{} energy. 
The comparison between neutron emission probabilities of Ref.~\cite{Mumpower2016} and those derived from the current work's spectra provide another quantification of uncertainty on the \bstr{} smoothing parameter. 

The normalization constant, $C_\omega$, of Eqn. \ref{eq:omega} is given by the condition
\begin{equation}
 \int_{0} ^{Q_\beta}\omega(E_x) \textrm{d}E_x = 1 ,
  \label{eq:omega_norm}
\end{equation}
where $Q_{\beta}$ is the \bmd{} energy window. 

\begin{table}[h]\label{table:dev}
\centering
\caption{The average maximum deviation in multiplicity for particle channel $x$, $N_{x}$, and average energy, $\langle E\rangle$, of our \bmd{} spectra for the individual particles based on the change in the \bstr{} smoothing parameter, $\Omega$. }
\begin{tabular}{c|c|c}
\textbf{Particle} & \textbf{Avg. Max Deviation in $N_{x}$} & \textbf{Avg. Max Deviation in $\langle E\rangle$} \\ \hline
Gamma             & 0.225                                      & 0.088                                \\ \hline
Neutron           & 0.274                                      & 0.119                                \\ \hline
Electron          & 0                                          & 0.556                                \\ \hline
Neutrino          & 0                                          & 0.660                                \\ \hline
\end{tabular}
\end{table}

The second component of the QRPA+HF model involves tracking the statistical decay of the daughter nucleus, a process governed by the Hauser-Feshbach formalism. 
This approach is rooted in the Bohr independence hypothesis of compound nucleus formation, which asserts that the formation and subsequent decay of the compound nucleus are independent processes, a concept originally formulated within the statistical model of nuclear reactions \cite{Weisskopf1956}. 
This assumption can be understood as a manifestation of ergodicity in nuclear dynamics, where the internal degrees of freedom of the nucleus equilibrate over timescales much shorter than the decay process, effectively “erasing” any memory of the entrance channel.

In this framework, the nucleus formed after an initial excitation redistributes its energy among a dense set of nuclear states before decaying. 
The probability of decay into a particular final state is then governed by the relative densities of accessible states and the associated transmission coefficients for the available reaction channels. 
This statistical treatment is particularly valid when the level density of the compound nucleus is sufficiently high, allowing for a quasi-continuum of states in which transitions occur according to well-defined statistical weights.

At high excitation energies, where experimental level information becomes sparse or unavailable due to the overwhelming density of nuclear states, a level density model must be employed. 
In our implementation, we utilize the Gilbert-Cameron level density model, which effectively bridges the low-energy discrete level region with the high-energy statistical continuum. 
This model consists of two primary components: a constant-temperature level density at low excitation energies and a Fermi gas description at higher energies. 
The transition between these two regimes ensures a smooth interpolation between the known discrete levels and the statistical continuum \cite{Kawano2006}. 

In our Hauser-Feshbach framework, the competition between neutron emission and gamma decay is treated using a statistical approach, where the probability of each decay channel is determined by the corresponding transmission coefficients and level densities \cite{Mumpower2018, Mumpower2022}. 
The statistical model accounts for the branching ratios between these competing processes, ensuring that the total decay probability sums to unity. 
Neutron emission is dominant when the excitation energy of the compound nucleus exceeds the neutron separation energy, while gamma decay becomes more relevant at lower excitation energies where neutron emission is energetically suppressed. 
For further details of the statistical HF model and relevant inputs see Ref.~\cite{Mumpower2016}. 
Next, we focus on the calculation of the particle spectra. 

\section{Calculation of Particle Spectra}
\label{sec:pspec}

In \bmd{}, an electron and an anti-neutrino are emitted when the nucleus undergoes transmutation from parent to daughter. 
The spectrum of these particles can be calculated via Fermi's theory of $\beta$-decay \cite{Fermi1934}. 
Following standard practice, the nucleus is assumed to carry away zero kinetic energy but arbitrary momentum (justified by its large relative mass), so that the total decay energy is divided between the electron and anti-neutrino and their momenta are uncorrelated.
For a given electron energy and neutrino neutrino energy, the momenta each have two degrees of freedom from the vector direction; this leads the density of final states in Eqn.~\ref{eq:fermi_gr} to take the form:
\begin{equation}
    \rho(Q_\beta) \sim p_e^2 p_\nu^2 \delta (Q_\beta - E_e - E_\nu).
\end{equation}
The interaction between the nucleus and outgoing electron distorts the electron wave function (introducing the factor $F$ from Eq. 3), and in high-density environments, ambient electrons block otherwise available outgoing electron states (introducing a factor of $1-f(E_e)$, defined presently in Eq. \ref{eqn:e_block}).
Incorporating these elements, we have
\begin{equation}
    \rho(Q_\beta) \sim p_e^2 p_\nu^2 F(Z+1, A, E_e)(1-f(E_e)) \delta (Q_\beta - E_e - E_\nu).
\end{equation}
Using $p^2 = (E + m)^2 - m^2$, defining $\mathcal{E} \equiv E + m$, $\mathcal{Q} \equiv Q_\beta + m_e$, and treating the neutrino as massless, we arrive at
\begin{equation}
  \rho(Q_\beta) \sim (\mathcal{E}_e^2-m_e^2)\mathcal{E}_\nu^2 F(Z+1, A, \mathcal{E}_e)(1-f(\mathcal{E}_e)) \delta (\mathcal{Q} - \mathcal{E}_e - \mathcal{E}_\nu).
  \label{eq:lepton_dos}
\end{equation}
From here, we are positioned to construct the shapes of the electron and neutrino energy spectra.
To obtain the spectrum $S$ of either lepton, integrate \ref{eq:lepton_dos} over the energy of the other lepton; this eliminates the $\delta$ function.
\begin{align}
    S_e(\mathcal{E}_e) &\sim (\mathcal{E}_e^2 - m_e^2) (\mathcal{Q} - \mathcal{E}_e)^2 F(Z+1, A, \mathcal{E}_e)(1-f(\mathcal{E}_e)) \label{eq:e_spectrum_unscaled} \\
    S_\nu(\mathcal{E}_\nu) &\sim ((\mathcal{Q} - \mathcal{E}_\nu)^2 - m_e^2) \mathcal{E}_\nu^2 F(Z+1, A, \mathcal{Q} - \mathcal{E}_\nu)(1-f(\mathcal{Q} - \mathcal{E}_\nu))
    \label{eq:nu_spectrum_unscaled}
\end{align}
Integrating \ref{eq:e_spectrum_unscaled} over electron energy gives the phase-space integral, we denote as $g$ below, typically used in $\beta$-decay calculations. 
For the present purpose, however, all that remains is to reintroduce the factors that scale the spectra and put them in the correct units, then sum over final states in the daughter nucleus.
\begin{align}
    S_{e,i}(\mathcal{E}_e) &= \frac{\lambda_0}{m_e^5} \sum_f B_{if}  (\mathcal{E}_e^2 - m_e^2) (\mathcal{Q}_{if} - \mathcal{E}_e)^2 F(Z+1, A, \mathcal{E}_e)(1-f(\mathcal{E}_e)) \label{eq:e_spectrum} \\
    S_{\nu,i}(\mathcal{E}_\nu) &= \frac{\lambda_0}{m_e^5} \sum_f B_{if} ((\mathcal{Q} - \mathcal{E}_\nu)^2 - m_e^2) \mathcal{E}_\nu^2 F(Z+1, A, \mathcal{Q} - \mathcal{E}_\nu)(1-f(\mathcal{Q} - \mathcal{E}_\nu))
    \label{eq:nu_spectrum}
\end{align}
In these equations, $i$ and $f$ index the initial and final nuclear states.
$\lambda_0$ contains physical constants, and for the decays considered in this work, $\lambda_0 \approx \frac{\ln 2}{10^{3.596}}$ s$^{-1}$ \cite{brown1978empirical}.
The factor of $B_{if}$ is the reduced square nuclear transition matrix element and, with dimensionful constants absorbed by $\lambda_0$, plays the role of $\vert \langle f \vert \widehat{O}_\beta \vert i \rangle \vert^2$ from Eqn. \ref{eq:fermi_gr}.
The blocking factor $1-f(E)$ is simply Pauli blocking in a Fermi-Dirac distribution of ambient electrons.
\begin{equation}
  f(E_e) = \frac{1}{1+e^{\frac{E_e-\mu_e}{T}}} 
  \label{eqn:e_block}
\end{equation}
Here, $\mu_e$ is the electron chemical potential and $T$ is the temperature in units of energy. 
Since we reserve ourselves to the study of decays from the ground-state (GS) of the parent nucleus, $i$ = GS in Eqs. \ref{eq:e_spectrum} and \ref{eq:nu_spectrum}, $T=0$, and all outgoing electron states are available $\rightarrow f(E_e) = 0$. 

Subsequent particle spectra (of the emitted neutrons and \grays{}) is calculated by following the statistical de-excitation in the Hauser-Feshbach code. 
We denote the compound state as $c^{(j)}_k$ where $c^{(j)}$ represents the compound nucleus after emitting $j$ neutrons, and $k$ indexes the excited states in this nucleus starting from the ground state ($k=0$) and proceeding higher. 
Using this definition, $c^{(0)}$ would be shorthand for describing the $\beta$-decay daughter nucleus ($Z+1$, $A$) and $c^{(j)}$ would describe ($Z+1$, $A-j$). 

Label the discrete energy levels of the $c^{(j)}$ compound nucleus by $E_1$, $E_2$, $\ldots$, $E_i$, $\ldots$, $E_N$, with the energies ordered, $E_1 < E_2 < \cdots < E_N$. 
Energies in the neutron-emission daughter nucleus (with $j>0$) are marked with primes. 
The initial population of the levels from the $\beta$-decaying parent is defined by the strength function and the phase space factor,
\begin{equation}
 P_0(E_i) = B S_\beta(E_i) g(Q_\beta - E_i) \ ,
 \label{eqn:daughter_p0}
\end{equation}
where $B$ is a normalization factor that is defined by integrating over all energies $E_i$ in the daughter nucleus, and $g$ is the phase space factor that accounts for the available energy of the leptons. 

The unnormalized $\gamma$-emission transition probability from state $i$ to state $k$ in the $j$-th compound nucleus is defined as
\begin{equation}
 p^{(j)}(E_i \rightarrow E_k) \propto T^{(j)}_\gamma(E_i - E_k) \rho^{(j)}(E_k)
 \label{eqn:pprob}
\end{equation}
where the transition is from a level with high excitation energy $E_i$ to a level of energy $E_k$, $T^{(j)}_\gamma$ is the $\gamma$-ray transmission coefficient for $c^{(j)}$, $\rho^{(j)}(E_k)$ is the level density in $c^{(j)}$ at energy $E_k$, and $N^{(j)}(E_i)$ is a normalization factor defined below. 
The neutron-emission transition probability from state $i$ in $c^{(j)}$ to state $k'$ in $c^{(j+1)}$ is defined as
\begin{equation}
 q^{(j)}(E_i \rightarrow E_{k^\prime}) \propto T^{(j+1)}_n(E_i-S^{(j)}_n-E_{k^\prime}) \rho^{(j+1)}(E_{k^\prime})
 \label{eqn:qprob}
\end{equation}
where $S^{(j)}_n$ is the neutron separation energy of $c^{(j)}$, $T^{(j+1)}_n$ is the neutron transmission coefficient from $c^{(j+1)}$ to $c^{(j)}$, and $\rho^{(j+1)}(E_{k^\prime})$ is the level density in $c^{(j+1)}$ evaluated at $E_{k^\prime}$. 
Recall that the prime on the $k$ index indicates that the energy level is in a different compound nucleus. 

To calculate branching ratios from state $E_i$, we must normalize by summing over all possible exit channels. 
In this case, the exit channels are $\gamma$ and neutron emission. 
\begin{equation}
 N_j(E_i) = \sum_{E_i> E_k} T^{(j)}_{\gamma}(E_i-E_k) \rho^{(j)}(E_k)
          + 
          \sum_{E_i>E_k\prime} T^{(j+1)}_{n}(E_i-S^{(j)}_n-E_{k^\prime}) \rho^{(j+1)}(E_{k^\prime})
 \label{eqn:normprob}
\end{equation}
The first sum represents $\gamma$ transitions to lower states in the same nucleus. 
The second sum represents neutron emission leading to levels $E_{k{\prime}}$ in a subsequent daughter nucleus. 
The branching ratio for $\gamma$ emission from energy level $E_i$ is 
\begin{equation}
  B_{\gamma}(E_i \to E_k) = \frac{p^{(j)}(E_i \to E_k)}{N_j(E_i)} \ ,
 \label{eqn:brat_gam}
\end{equation}
and the branching ratio for neutron emission from the same level is 
\begin{equation}
  B_{n}(E_i \to E_{k^\prime}) = \frac{q^{(j)}(E_i \to E_{k^\prime})}{N_j(E_i)} \ .
 \label{eqn:brat_n}
\end{equation}
The branching ratios are unitless probabilities, while $p$, $q$ and $N$ have units of [1/MeV]. 

With these terms defined, we may now determine the probability $P(E_x)$ of landing in a state $E_x$ in the daughter nucleus for the case of no neutron emission ($j=0$), 
\begin{equation}
  P_{(j=0)}(E_x) = P_0(E_x) + \sum_{i > x} P(E_i) B_{\gamma}(E_i \to E_x) \ .
 \label{eqn:prob_daugter_j0}
\end{equation}
In this equation, the first term represents direct population from $\beta$-decay and the second term takes into account $\gamma$ feeding from higher-lying levels within the same nucleus. 
If any neutron emission occurs from states above the neutron threshold, this reduction in level population is accounted for via the normalization factor. 
Once the daughter nucleus is defined by this population, one must then calculate the nuclear de-excitation, taking into account competition between neutron emission and photon emission. 

The probability to populate a level in the neutron-emission daughter, compound system $j+1$, is 
\begin{equation}
  P_{(j=1)}(E_{x^\prime}) = \sum_{i} P(E_i) B_n(E_i \to E_{x^\prime}) + \sum_{k^{\prime} > x} P(E_{k^{\prime}}) B_{\gamma}(E_{k^{\prime}} \to E_{x^\prime}) \ .  
 \label{eqn:prob_daugter_j1}
\end{equation}
The first term takes into account direct population via neutron emission from the preceding nucleus while the second term takes into account $\gamma$ transitions from higher-lying states in the daughter nucleus. 
Note the recursion in the last two equations. 
A given $P(E_i)$ must also again use the same equation to find its respective feeding until the highest accessible level $k$ can only be directly populated via $\bar{S}_{\beta}(E_k)$. 
This type of function can be handled naturally on a computer. 

Without loss of generality the equations can be extended to $j$ neutron emission along the initial $\beta$-decaying daughter's isotopic chain 
\begin{equation}
P(E_x) =
\begin{cases}
P_{(j=0)}(E_x), & \text{if } E_x \text{ is in the first daughter nucleus}, \\
P_{(j=1)}(E_x), & \text{if } E_x \text{ is in the second daughter nucleus}, \\
\vdots & \vdots 
\end{cases}
 \label{eqn:PofE}
\end{equation}
The equations may also be generalized to include more exit channels, for example with the emission of other charged particles. 
Since we are primarily focused on neutron-rich nuclei, this latter generalization is not considered in this work. 

For neutron-rich nuclei we can compute the $j$-th neutron emission probability ($j\textrm{n}$) by considering all the channels that end in the $j$-th compound nucleus relative to all possible pathways:
\begin{equation}
 P_{j\textrm{n}} = P_j(E_{\textrm{gs}})
 \label{eqn:Pjn}
\end{equation}
where we evaluate the level population at the ground state energy, $E_{\textrm{gs}}$, in $c^{(j)}$. 
In general a $P_{j\textrm{n}}$ value may increase or decrease relative to calculations based on the energy window argument that has been used in past work. 
Beta-decay with no neutron emission is denoted as $P_{0\textrm{n}}$ (just Equation \ref{eqn:prob_daugter_j0}) and because of the normalization in the calculation of the level population we have a similar constraint on the neutron emission probabilities, $\sum_{j=0} P_{j\textrm{n}} = 1$. 

We use the above probabilities to calculate the $\gamma$ emission spectra. 
The $\gamma$ spectra per unit energy are
\begin{equation}
 \frac{\textrm{d}N_\gamma}{\textrm{d}E}(E_{\gamma}) = \sum_{j} \sum_{i} P_{j}(E_i) \sum_{k < i} \frac{B_{\gamma}^{j}(E_i \to E_k)}{\Delta E_{\gamma}} \Theta(E_{\gamma} - (E_i - E_k)) \ ,
 \label{eqn:spec_gamma}
\end{equation}
where $E_{\gamma}$ is the $\gamma$-ray energy, $j$ indexes all the daughter nuclei, $P_{j}(E_i)$ is the population probability of level $E_i$ in nucleus $j$, $B_{\gamma}^{j}(E_i \to E_k)$ is the $\gamma$ branching ratio for transitions within nucleus $j$, $\Theta(E_{\gamma} - (E_j - E_i))$ is a step function to ensure $\gamma$ emission occurs at the expected energy, and $\Delta E_{\gamma}$ is the energy bin width. 
Delayed neutron spectra have been covered in the context of the QRPA+HF model in a previous study \cite{Kawano2008}. 
Here we provide the equation used to calculate the spectra, 
\begin{equation}
\frac{\textrm{d}N_n}{\textrm{d}E}(E_n) = \sum_{j} \sum_{i} P_{j}(E_i) \sum_{k^{\prime}} \frac{B_n^{j}(E_i \to E_{k^{\prime}})}{\Delta E_n} \Theta(E_n - (E_i - S_n^{j} - E_{k^{\prime}})) \ ,
 \label{eqn:spec_neutrons}
\end{equation}
where $E_n$ is the neutron energy, $P_{j}(E_i)$ is the population probability of level $E_i$ in nucleus $j$, $B_n^{j}(E_i \to E_{k^{\prime}})$ is the neutron branching ratio between the two states, $\Theta(E_n - (E_i - S_n^{j} - E_{k^{\prime}}))$ is a step function to ensure neutron emission occurs at the expected energy, and $\Delta E_{n}$ is the energy bin width. 
The units of the $\frac{\textrm{d}N}{\textrm{d}E}$ quantities are in MeV$^{-1}$ such that the integration over energy yields the particle multiplicity for photons and neutrons respectively,
\begin{align}
  N_\gamma &= \int_{0}^{Q_{\beta}} \frac{dN_\gamma}{dE} \, dE_\gamma \ , \label{eqn:gspec} \\
  N_n &= \int_{0}^{Q_{\beta}} \frac{dN_n}{dE} \, dE_n \ .
  \label{eqn:nspec}
\end{align}
The average energy of photons and neutrons can be calculated by computing the energy weighted integrals
\begin{align}
  \langle E_\gamma \rangle &= \int_{0}^{Q_{\beta}} E_\gamma \frac{dN_\gamma}{dE}  \, dE_\gamma \ , \label{eqn:Eavg_g} \\
  \langle E_n \rangle &= \int_{0}^{Q_{\beta}} E_n \frac{dN_n}{dE} \, dE_n \ .
  \label{eqn:Eavg_n}
\end{align}

De-excitation of a nucleus from a discrete level may also include internal conversion (IC). 
Internal conversion is a process where an excited nucleus transfers its excess energy directly to one of its own atomic electrons that is then ejected from the atom. 
The atom then fills the vacancy left by the first electron from a higher lying atomic orbital, emitting characteristic X-rays or Auger electrons. 
This process may compete with $\gamma$-ray emission, becoming more probable in nuclei where the $\gamma$-ray transition energy is low or the atomic number is high. 
Auger electrons can also be emitted after photoionization by a nuclear $\gamma$-ray. 
The ratio of the number of conversion electrons to the number of photons emitted is the internal conversion coefficient ($\alpha$). 
In nuclear transitions such as $0^{+} \rightarrow 0^{+}$ this is the primary decay process. 

We include measured internal conversion rates in the calculation of our $\gamma$-ray spectra. 
The impact of this inclusion is minor as $\alpha$ values are typically not available. 
Values of $\alpha$ are taken from relevant ENSDF files when available. 
In the case that no information is available, we use $\alpha = 0$. 
See Section \ref{sec:nonnuc} for more information regarding atomic effects. 

When calculating the $\beta$-decay particle spectra, $Q_\beta$ is an important input quantity that defines the maximum energy available in the decay. 
Neutron separation energies also define the maximum energy available for emitted neutrons. 
We use the FRDM12 masses when one or more of the participating species's mass is unknown. 
When both the parent and daughter masses are measured, the AME20 is used for $Q_\beta$ \cite{Wang2021}. 
It is crucial to note that theoretical predictions and evaluated data for masses are not mixed. 
If they were mixed, prominent shifts can arise from the mismatch of model and data (generally on the order of the $\sigma_\textrm{RMS}$ of the mass model, which is typically around 500 keV or more).
Both $Q_\beta$ or the $S_{1\textrm{n}}$ values can be impacted in this regard, causing subsequent \textit{unphysical} alterations in the predicted particle spectra. 
The variation of our results with other theoretical mass predictions are available upon request. 

\section{Model Output and Sensitivities} \label{sec:output}
\begin{figure}
\begin{centering}
\includegraphics[width=85mm]{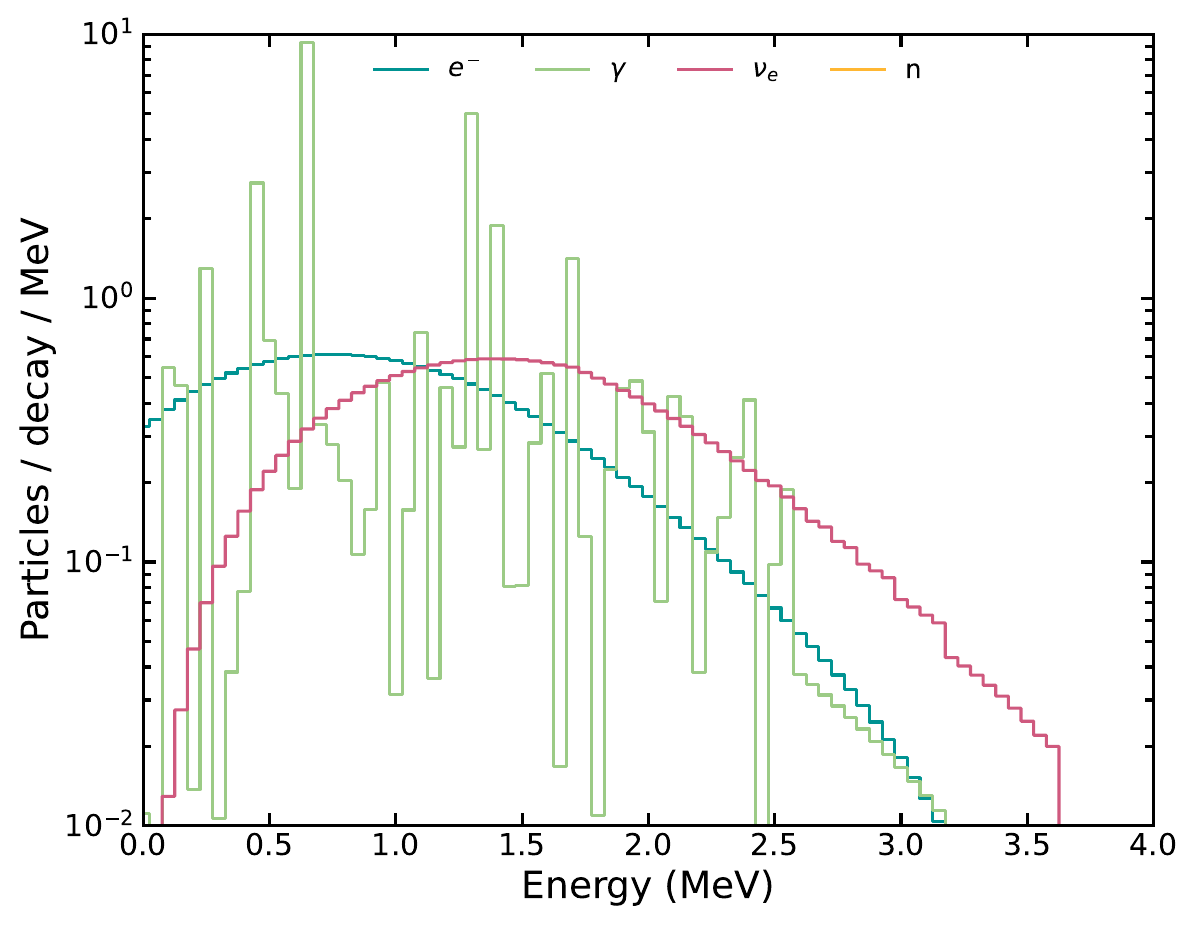}
\includegraphics[width=85mm]{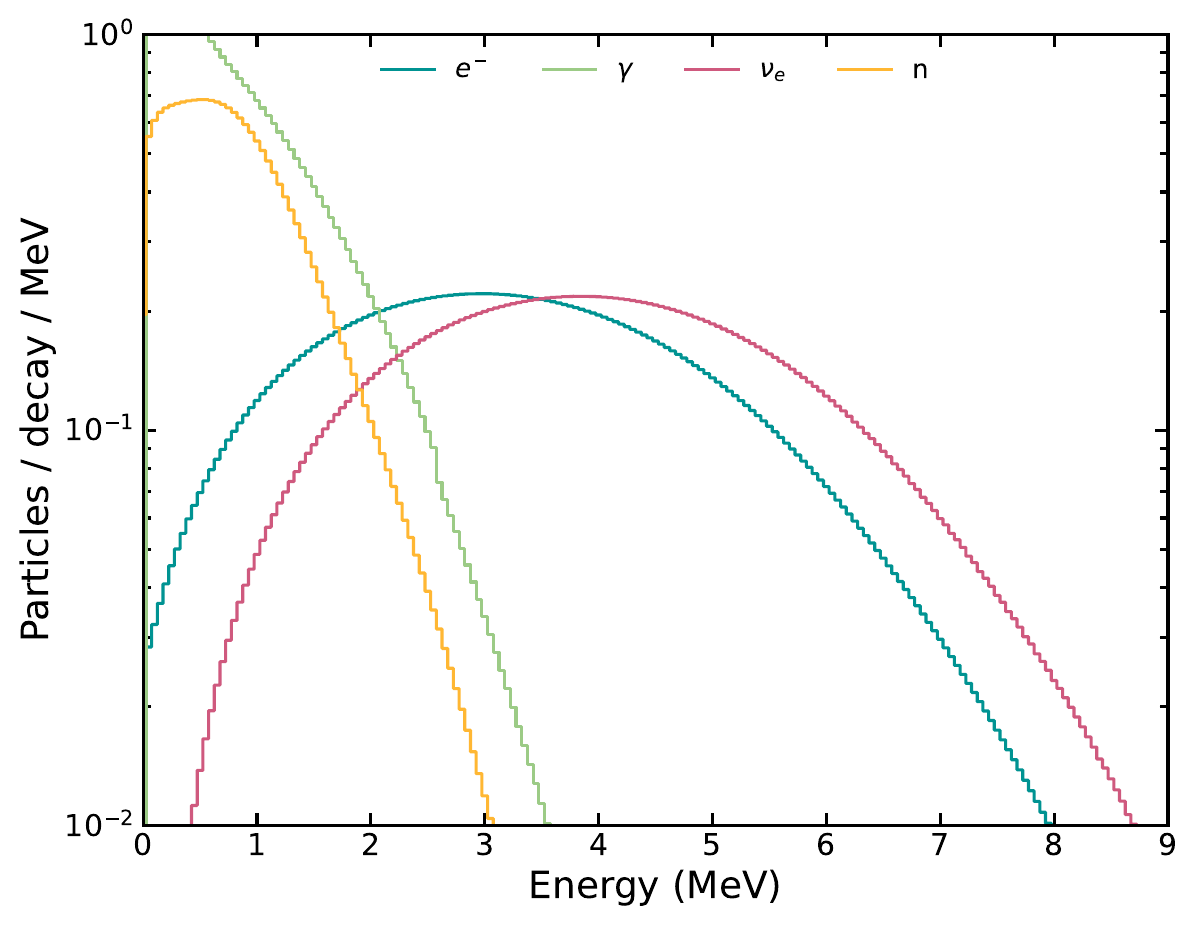}
\caption{The \bmd{} particle energy spectra of (left) $^{139}$Cs and (right) $^{155}$Cs. In the left panel the nucleus' decay daughter has many known levels resulting in discrete $\gamma$ lines while in the right panel, $^{155}$Cs is predicted to emit a delayed neutron. }
\label{fig:spec}
\end{centering}
\end{figure}

Two example particle spectra from the $\beta$-decay of $^{139}$Cs and $^{155}$Cs ($Z=55$) are shown in the left and right panels of Figure \ref{fig:spec} respectively. 
Emitted particles includes electrons ($e^{-}$ or $\beta$), $\gamma$-rays, neutrinos ($\nu$), and neutrons (n). 
The decay of $^{139}$Cs populates $^{139}$Ba with over 100 known levels below $Q_\beta$. 
The $\gamma$ spectra for this decay is therefore less smooth than in the right panel where no data is known. 
For $^{155}$Cs, the multiplicity of $\gamma$-rays is $N_\gamma=3.33$ and for neutrons $N_n=0.024$ for this decay. 
Average energies for particle emission are $\langle E_\gamma \rangle=0.703$ MeV, $\langle E_n \rangle=0.4130$ MeV, $\langle E_\beta \rangle=2.998$ MeV, and $\langle E_\nu \rangle = 3.751$ MeV. 
These values can be confirmed by observing the peaks in the distributions shown in Figure \ref{fig:spec}. 
For this decay, the FRDM12 predicted Q value is $Q_\beta=10.25$ MeV. 
Extrapolation from the AME20, the $Q$ value is instead $Q_\beta=7.64$ MeV, highlighting an intrinsic uncertainty that exists in our calculations for neutron-rich nuclei: unmeasured ground state masses. 

\begin{figure}
\begin{centering}
\includegraphics[width=120mm]{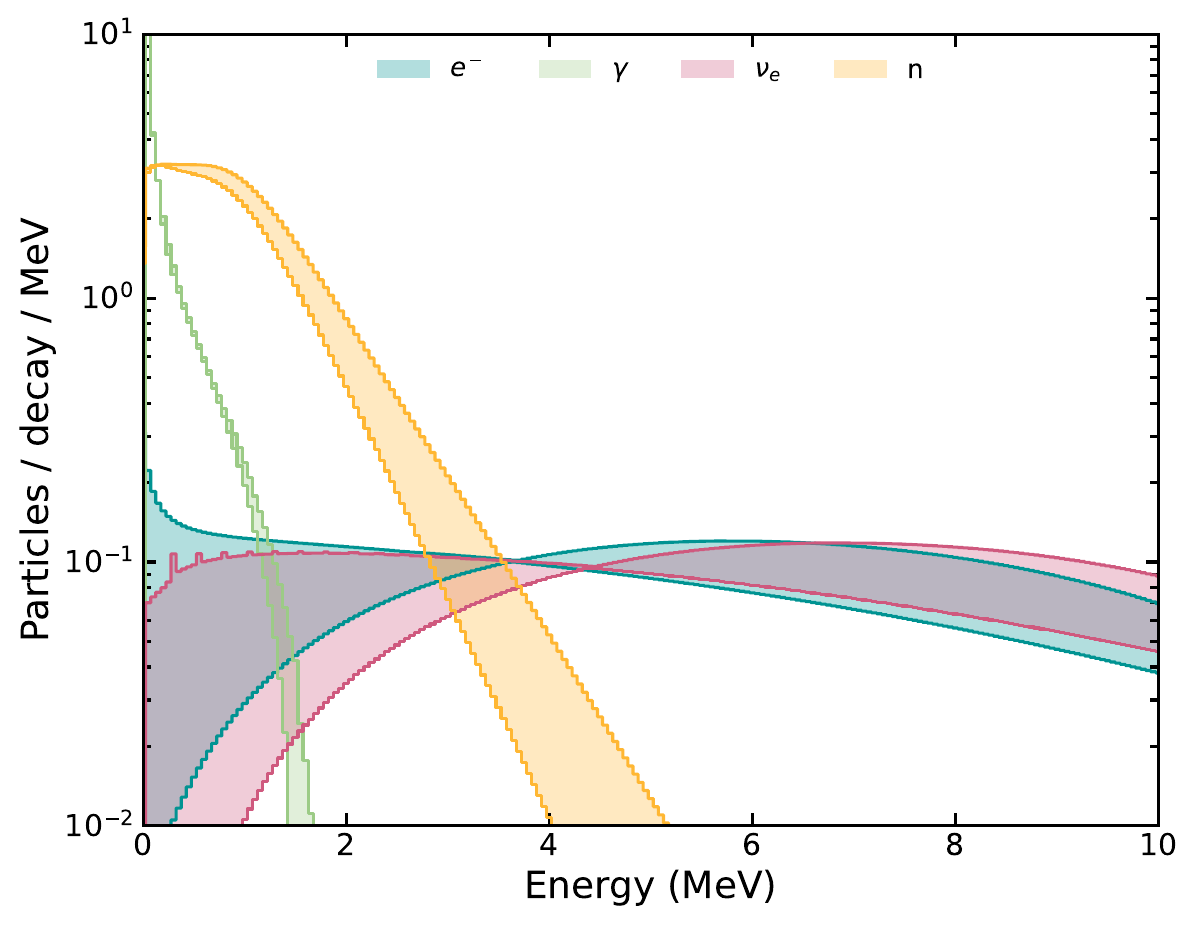}
\caption{The spread in \bmd{} particle energy spectra of the unmeasured nucleus $^{165}$Sn for three choices of $\Omega$. This nucleus is predicted to be a delayed neutron emitter. }
\label{fig:spec-dev}
\end{centering}
\end{figure}

The calculated particle spectra are also sensitive to the $\beta$-strength distribution from QRPA. 
To simulate this uncertainty we run three smoothing procedures for the $\beta$-strength, $\Omega \in \{30.0, 100.0, \Omega(A)\}$. 
As mentioned previously, this uncertainty can impact the multiplicity of $\gamma$'s and neutrons, however electrons and neutrinos remain unaltered as there is always just a single particle emitted of each respectively. 
The difference in the particle energy distributions can be quite large for the varying choices of $\Omega$, as shown in Figure \ref{fig:spec-dev}. 
This figure reinforces the importance of making accurate $\beta$-strength predictions, which in turn relies on the prediction of nuclear structure for exotic nuclei.  

\begin{figure}
\begin{centering}
\includegraphics[width=140mm]{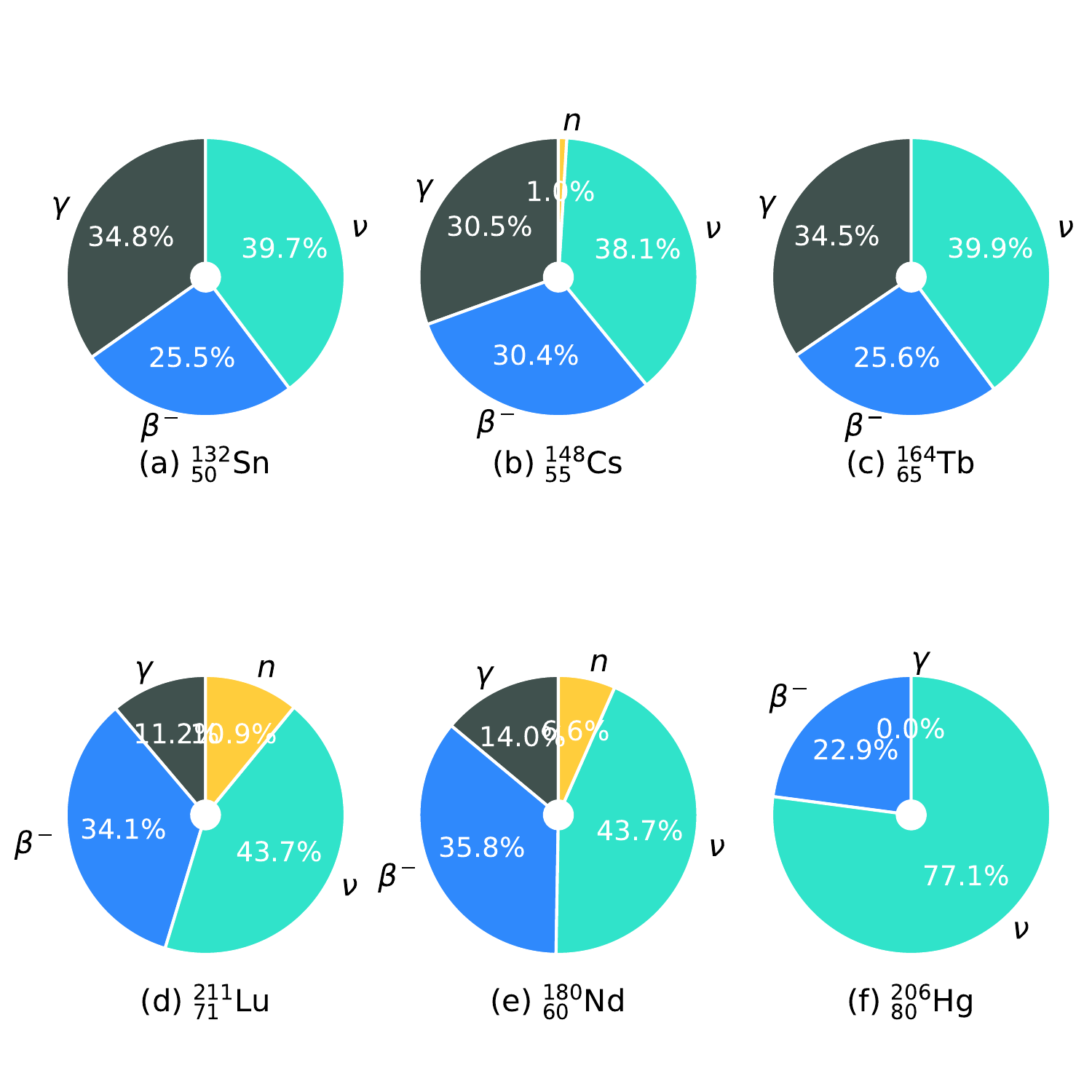}
\caption{The energy partition between electrons, \grays{}, neutrinos and neutrons associated with the \bmd{} of select nuclei.}
\label{fig:Epart}
\end{centering}
\end{figure}

Energy partitioning in $\beta$-decay has important consequences for a host of applications. 
The total energy release from the emitted particles is
\begin{equation}
 E_\textrm{release} = N_\gamma \langle E_\gamma \rangle + N_n \langle E_n \rangle + N_\beta \langle E_\beta \rangle + N_\nu \langle E_\nu \rangle \ ,
 \label{eqn:totE}
\end{equation}
where this quantity is less than or equal to $Q_\beta$ for a typical decay. 
In $\beta$-decay, $N_\beta = N_\nu = 1$. 
The partition of a specified particle, $x$, is defined by the ratio
\begin{equation}
  G_x = \frac{N_x \langle E_x \rangle}{E_\textrm{release}} \ .
  \label{eqn:Epart}
\end{equation}
We show the energy partition for select nuclei in Figure \ref{fig:Epart}. 
We find that electrons typically range from 20\% to 35\% of the energy release.
Neutrinos typically take around 40\% of the energy, but can take a higher percentage for $\beta$-decaying nuclei that reside closer to stability. 
Gamma-rays and neutrons take up the remainder of the energy reservoir. 
For nuclei near stability, $Q_\beta < S_{1\textrm{n}}$ and no neutrons are released in $\beta$-decay; the multiplicity and average energy is zero. 

Across the chart of nuclides, for most nuclei, the bulk of the energy in \bmd{} is taken away by the neutrino. 
However, there are a few exceptions. 
Some spherical and deformed nuclei have $\gamma$-rays as the most dominant fraction of the total energy release. 
In extremely neutron-rich nuclei beyond the neutron dripline, neutrons may take away the highest fraction of the energy released due to a unique feature of their production. 

For almost every nuclear $\beta$-decay across the chart of nuclides, $E_\textrm{release} < Q_\beta$, as one would expect via conservation of energy. 
However, the case can arise that $E_\textrm{release} > Q_\beta$ while still conserving energy. 
When nuclei beyond the neutron dripline participate in the decay process, the maximum excitation energy in one (or more) of the populated daughter nuclei may be greater than $Q_\beta$. 
Beyond the dripline, it is usually first that odd-$N$ nuclei that are unbound to neutron emission, $S_{1\textrm{n}} < 0$, and this additional energy is added to the available window. 
Even if some of the even-$N$ isotopes along the decay chain may have $S_{1\textrm{n}} > 0$ the scenario can still ensue so long as at least one nucleus in the decay chain is encountered with $S_{1\textrm{n}} < 0$. 
The increase to the decay energy can result in an increase to the mean emission energy of neutrons and $\gamma$'s.  
This situation is very sensitive to the masses of nuclei that remain unmeasured and thus this effect could vary between different model predictions. 
The effect may impact the study of nuclei that exist in the crusts of neutron stars. 

\section{Global Results}
\label{sec:results}

\begin{figure}
\begin{centering}
\includegraphics[width=\textwidth]{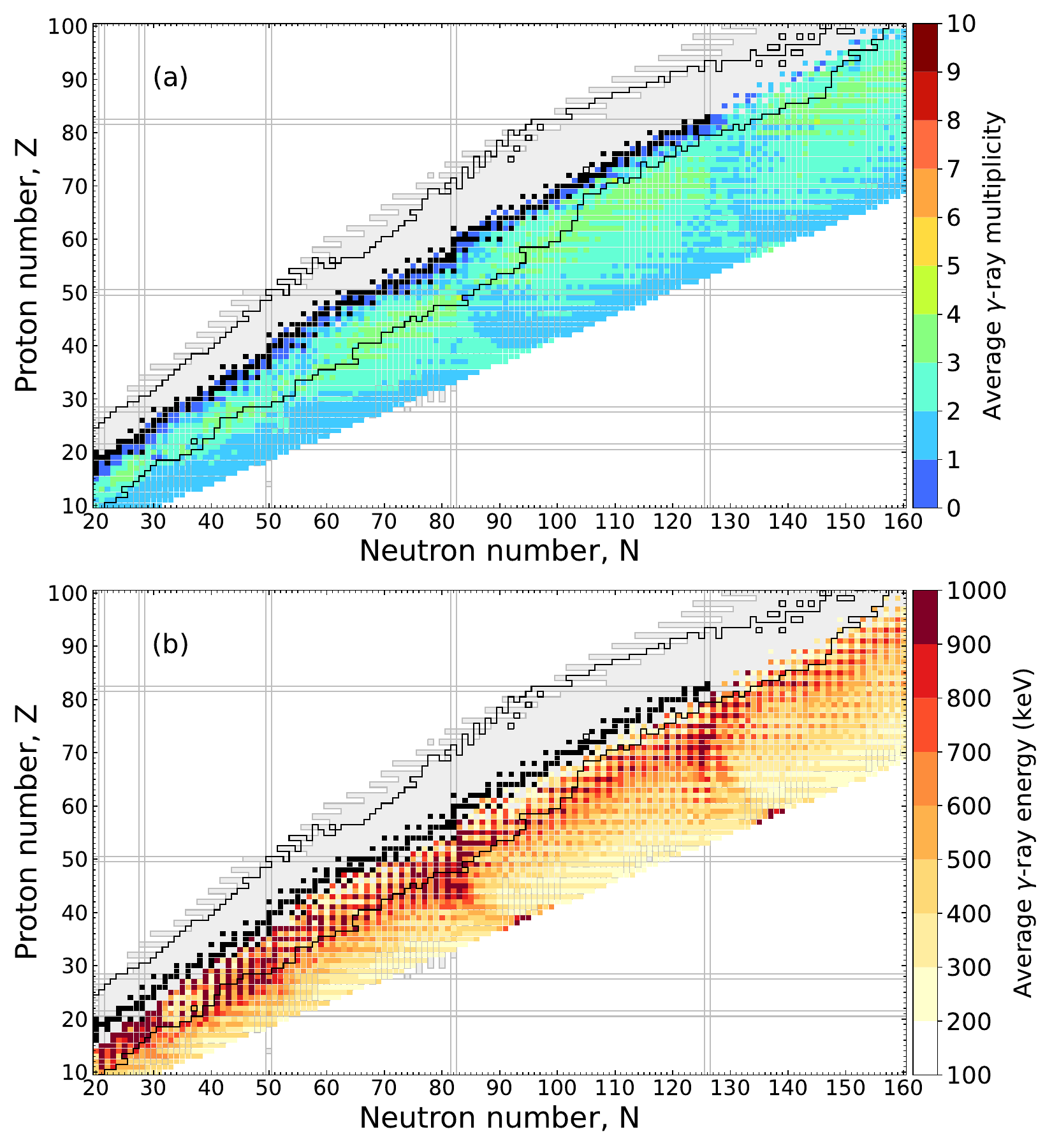}
\caption{(a) The \gray{} multiplicity and (b) average \gray{} energy after \bmd{} across the chart of nuclides. Black shading denotes \bstab{} nuclei and gray denotes the extent of bound nuclei in the FRDM12 model \cite{Moller2016}. Bounded region (black outline) indicates extent of measured masses part of AME20. }
\label{fig:avgG}
\end{centering}
\end{figure}

\begin{figure}
\begin{centering}
\includegraphics[width=\textwidth]{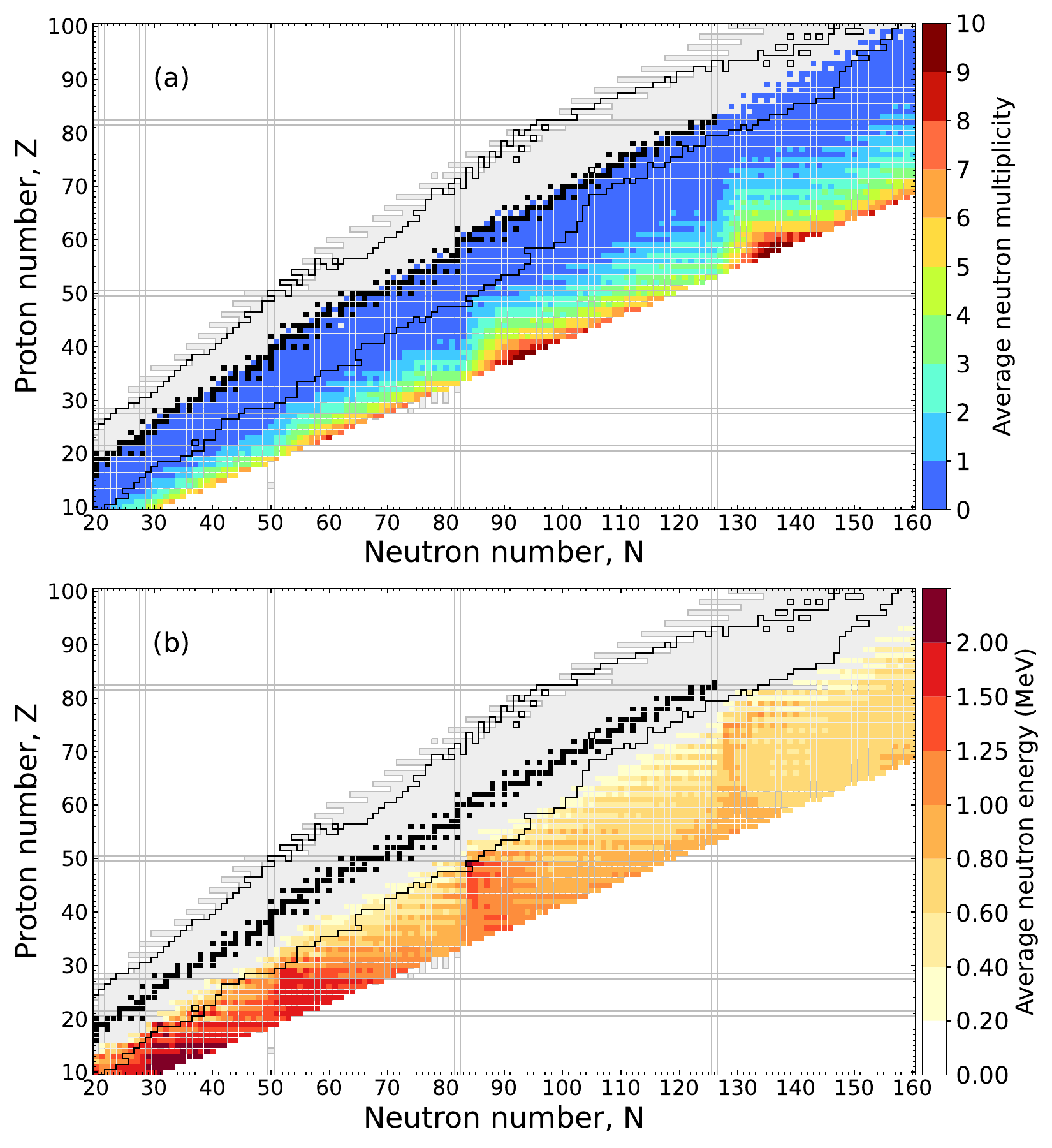}
\caption{(a) Neutron multiplicity and (b) average neutron energy after \bmd{} across the chart of nuclides. Black shading denotes \bstab{} nuclei and gray denotes the extent of bound nuclei in the FRDM12 model \cite{Moller2016}. Bounded region (black outline) indicates extent of measured masses part of AME20. }
\label{fig:avgN}
\end{centering}
\end{figure}

\begin{figure}
\begin{centering}
\includegraphics[width=\textwidth]{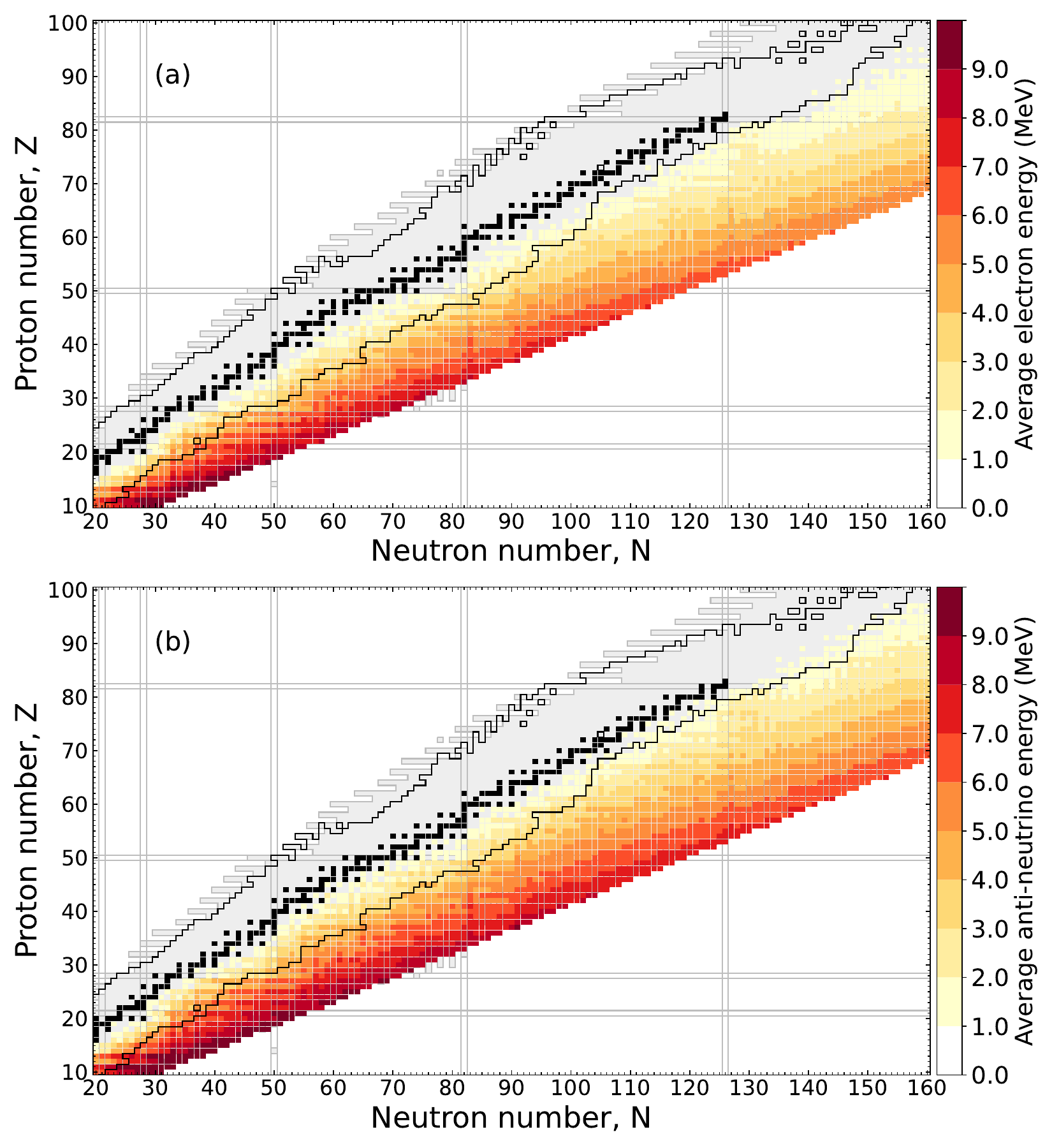}
\caption{(a) The average electron and (b) anti-neutrino energy after \bmd{} across the chart of nuclides. Black shading denotes \bstab{} nuclei and gray denotes the extent of bound nuclei in the FRDM12 model \cite{Moller2016}. Bounded region (black outline) indicates extent of measured masses part of AME20. }
\label{fig:avgNuB}
\end{centering}
\end{figure}

We summarize the particle multiplicity and average energy calculations across the chart of nuclides in Figures \ref{fig:avgG}, \ref{fig:avgN}, and \ref{fig:avgNuB}. 
In these calculations we have used AME20 masses when available, and fallen back to FRDM12 when no measurements exist. 
To indicate the use of AME20, we highlight the nuclei encompassed closer to stability with a bold black line.
The bulk of our calculations are predictions. 
We discuss trends below. 

Of all the particle spectra, $\gamma$-ray have the most structure in both as a function of energy and between neighboring nuclei. 
Gamma-ray multiplicity also deviates for neighboring nuclei depending sensitively on the structure of daughter nuclei. 
The average $\gamma$-ray energy is larger near closed shells, owing to the large level spacing in these nuclei. 
Odd-$N$ nuclei also tend to have higher average $\gamma$-ray energies due to the structure in daughter even-$N$ nuclei
Odd even observations. 
Average $\gamma$-ray energy trails off as a function of neutron excess; here the level structure is unknown but is predicted to be less dense. 

Figure \ref{fig:avgN} shows several noteworthy trends. 
Neutron emission after $\beta$-decay is small for the majority of nuclei (blue). 
It is only for very neutron-rich nuclei, where $S_{1\textrm{n}}$ is small, that the multiplicity increases significantly with nuclei near the dripline having $\langle N_n \rangle$ in the range of 3 to 5. 
On average, neutrons are emitted at higher energies for lighter nuclei, just beyond closed shells, and in general further from stability. 
Above mass number $A\sim140$, we find that there are very few neutron emitters with average neutron energies above 1 MeV. 

Electrons and neutrinos naturally have a multiplicity of unity. 
Therefore we only show the average energies in Figure \ref{fig:avgNuB}. 
The average energy for both particles increases somewhat monotonically with $Q_\beta$ and is relatively larger for lighter nuclei at the dripline in comparison to heavier nuclei. 
The electron energy plays a particularly important role in the electromagnetic follow up of astrophysical events \cite{Shenhar2024}. 

Above neutron number $N \sim 160$ and for $Z \gtrsim 90$ fission may follow $\beta$-decay in a process known as $\beta$-delayed fission \cite{Thielemann1983, Mumpower2018, Mumpower2022}. 
In this work we do not consider this mode as fission may take remarkable amount of the $\beta$-decay strength. 
Thus the fission channel will alter the number and energy of emitted particles when it participates. 

Our results are presented in the form of ASCII data tables associated with this work. 
Particle spectra files are named by the parent nucleus which undergoes \bmd{}: ``bspec\_Z\textbf{Z}A\textbf{A}.dat'' where \textbf{Z} is replaced by the parent nucleus proton number and \textbf{A} is the parent nucleus nucleon number. 
These files have several rows containing metadata followed by the data block consisting of the follow columns: Lower bound of energy bin (MeV), upper bound of energy bin (MeV), $\beta$ spectrum ($\frac{dN_\beta}{dE}$), $\nu$ spectrum ($\frac{dN_\nu}{dE}$), $\gamma$ spectrum  ($\frac{dN_\gamma}{dE}$), and neutron spectrum  ($\frac{dN_\textrm{n}}{dE}$). 
The spectra are calculated assuming the spreading of the $\beta$-strength obeys $\Omega(A)$ and have a constant energy binning of 50 keV. 
Recall that we have normalized the spectra such that the integration of the curve gives the particle multiplicity for \bmd{}.  

\section{Non-nuclear Electron Radiation}
\label{sec:nonnuc}

There are other types of electron radiation due to interactions of nuclear decay products with the surrounding atomic structure, namely Auger and Internal Conversion (IC) electrons. 
The theoretical calculation of these processes are not the subject of this study. 
However, we consider the experimental confirmed values here and tabulate their spectra because their impact may be important for various applications. 
The theoretical calculated spectrum from nuclear processes defined in Section \ref{sec:pspec} may be combined with the atomic processes outlined in this Section to form a more complete picture of the decay spectra for electrons. 

\begin{figure}
\begin{centering}
\begin{tabular}{c}
\includegraphics[width=\textwidth]{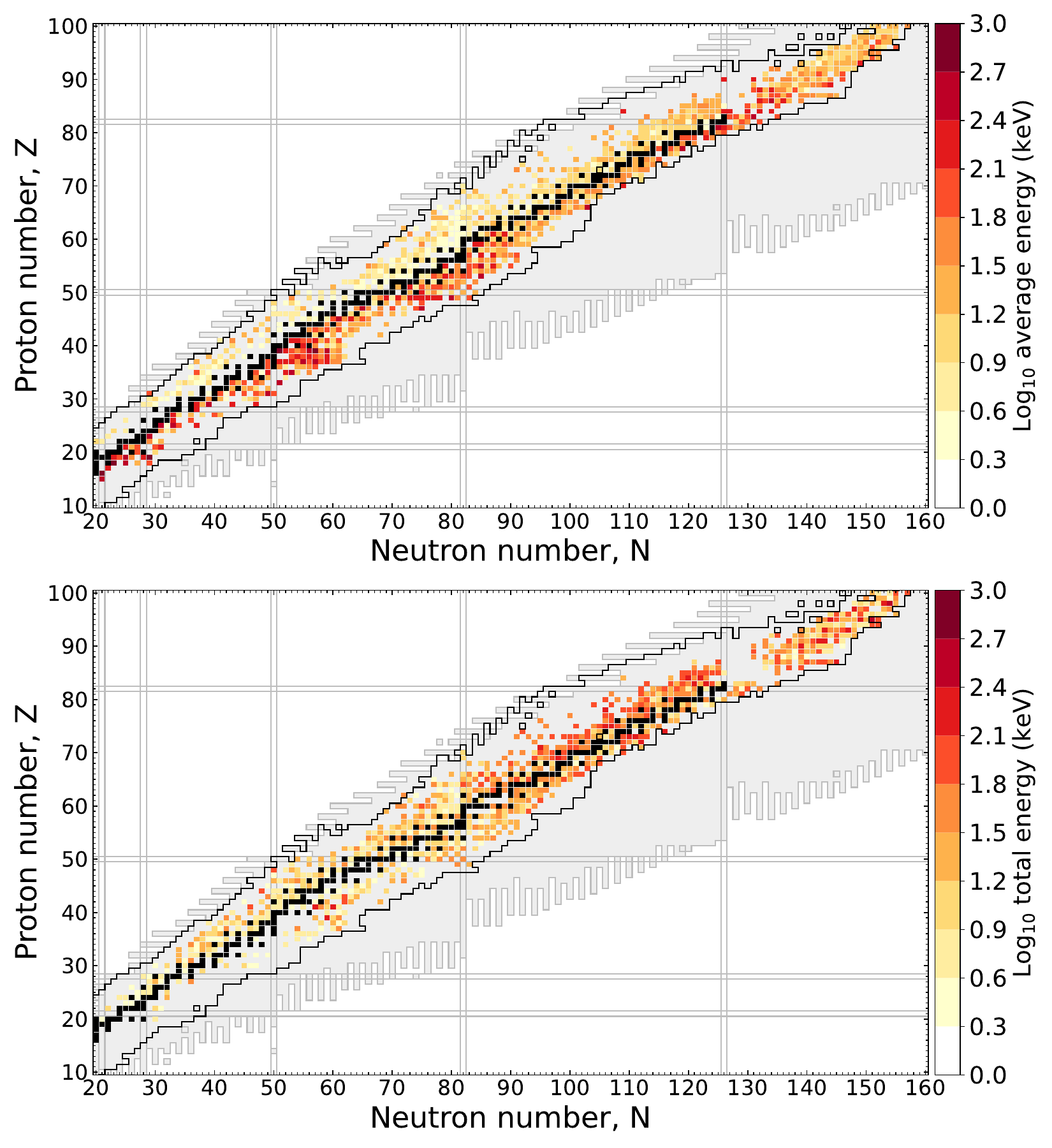}
\end{tabular}
\caption{(a) The average electron and (b) total electron energy for non-nuclear (Auger and Internal Conversion) electrons across the chart of nuclides, according to ENDF VIII.0 data tables~\cite{Brown2018}. Black shading denotes stable nuclei.}
\label{fig:auger_intconv}
\end{centering}
\end{figure}

Auger electron emission occurs when an inner-shell electron is removed, often due to photoionization, electron capture, or interaction with a $\beta$-particle. 
The vacancy left in the inner shell is filled by an electron from a higher-energy shell, and the excess energy is transferred to another (Auger) electron, which is then ejected from the atom. 
Internal conversion is a similar electron emission process triggered by the decay of an excited nucleus, when the excess nuclear energy is transferred to an orbital electron instead of emitting a $\gamma$ photon. 
When a nucleus transitions to a lower energy state, it can transfer its excess energy to an orbital electron instead of emitting a gamma photon. 
The electron, typically from the inner shells, is then ejected, and this is known as an IC electron. 
This process is more likely when the energy difference between nuclear states is relatively small, making gamma emission less probable.
Both Auger and Internal Conversion electrons can contribute significantly to the overall electron spectrum. 

We consider the average and total energy for non-nuclear electron emission across the nuclear chart and show the results in panels (a) and (b) of Figure~\ref{fig:auger_intconv}. 
This data is taken from the ENDF VIII.0 database~\cite{Brown2018}. 
Because of the trivial nature to parse this information, non-nuclear electron radiation is provided for both neutron-deficient and neutron-rich nuclei. 
As can be seen from the scale of the color bar, the contribution of atomic electrons are much smaller in energy than that from nuclear $\beta$-decay. 
These processes primarily alter the low-energy part of the electron spectrum. 
From the top panel, one can observe that the average electron energy is higher for neutron-rich nuclei than for neutron-deficient nuclei. 
The total energy from these non-nuclear processes increases as a function of mass number, as can be seen in the bottom panel of Figure~\ref{fig:auger_intconv}. 

ASCII tables associated with Auger and IC electron data are provided as supplemental material. 
These data files are labeled by the element symbol and nucleon number. 
Each data file has two columns. 
The first column gives the electron line energy (in keV) and the second column gives the intensity (counts per decay). 

\section{Acknowledgements}
We thank Peter M{\"o}ller for the use of the calculated $\beta$-decay strength data from \cite{Moller2019}. 
We thank Ani Aprahamian, Iris Dillmann, Jon Engel, George Fuller, Robert Grzywacz, Sean Liddick, Artemis Spyrou and Jin Wu for useful discussions on nuclear $\beta$-decay over the years. 
This work was supported by the US Department of Energy through the Los Alamos National Laboratory. 
Los Alamos National Laboratory is operated by Triad National Security, LLC, for the National Nuclear Security Administration of U.S. Department of Energy (Contract No. 89233218CNA000001). 

\section{Tabulated Data in Text} \label{sec:data}
In this section we provide a summary of the tabulated results, Table \ref{table:avg}, of the particle spectra associated with \bmd{}. 
The particle spectra themselves are not listed in the text as this data would take too many pages in length to describe. 
Consult the supplemental ASCII data tables for this information. 
In what follows we list for each \bmd{} the particle multiplicity and average energy and whether or not masses from the AME20 were used. 
A zero indicates that FRDM12 was used, and a 1 indicates that AME20 was used to define $Q_\beta$. 
This table is also provided in ASCII format as the file ``beta\_avg\_spec.dat'' in the associated supplemental material to the text. 
The first two lines of this file provides some metadata about the calculations. 
The third line of this file defines the column names and the remaining lines contain the data.

\begin{longtable}[c]{c|c|c|c|c|c|c|c|c|c}\label{table:avg}
\\
\caption{The average particle spectra quantities for the $\beta$-minus decay of the particular parent species.}

\textbf{$Z$} & \textbf{$N$} & \textbf{$A$} & $N_{\gamma}$ & $\langle E_\gamma \rangle$ & $N_{n}$ & $\langle E_n \rangle$ & $\langle E_e \rangle$ & $\langle E_{\nu} \rangle$ & Exp mass \\ \hline
\endfirsthead
\hline
\textbf{$Z$} & \textbf{$N$} & \textbf{$A$} & $N_{\gamma}$ & $\langle E_\gamma \rangle$ & $N_{n}$ & $\langle E_n \rangle$ & $\langle E_e \rangle$ & $\langle E_{\nu} \rangle$ & Exp mass \\ \hline
\endhead
\hline
\multicolumn{10}{|c|}{\textbf{Oxygen (O)}} \\
\hline
  8 &  11 &  19 & 0.00000e+00 & 0.00000e+00 & 0.00000e+00 & 0.00000e+00 & 1.54513e+00 & 1.94988e+00 & 1 \\ 
  8 &  12 &  20 & 0.00000e+00 & 0.00000e+00 & 0.00000e+00 & 0.00000e+00 & 6.24258e-01 & 8.64014e-01 & 1 \\ 
  8 &  13 &  21 & 1.50741e+00 & 1.35209e+00 & 0.00000e+00 & 0.00000e+00 & 2.82248e+00 & 3.27926e+00 & 1 \\ 
  8 &  14 &  22 & 2.34873e+00 & 1.10703e+00 & 1.85028e-02 & 2.60650e-01 & 1.69532e+00 & 2.12107e+00 & 1 \\ 
  8 &  15 &  23 & 3.14962e+00 & 1.32629e+00 & 5.19007e-02 & 5.85628e-01 & 3.15548e+00 & 3.64340e+00 & 1 \\ 
  8 &  16 &  24 & 1.94957e+00 & 8.15526e-01 & 3.62021e-01 & 9.45095e-01 & 3.58869e+00 & 4.08505e+00 & 1 \\ 
  8 &  17 &  25 & 2.39247e+00 & 5.28831e-01 & 6.37273e-02 & 1.76929e+00 & 6.93157e+00 & 7.46681e+00 & 1 \\
\end{longtable} 

\newpage

\bibliographystyle{unsrt}
\bibliography{refs.bib}

\end{document}